\documentclass[reprint,3p,times,onecolumn,10pt]{elsarticle}
\pdfoutput=1 
\usepackage{hyperref}
\usepackage{geometry}
\usepackage{mathtools}
\usepackage{graphicx}
\usepackage{latexsym}
\def\Dj{\hbox{D\kern-.73em\raise.30ex\hbox{-}
\raise-.30ex\hbox{}}}
\def\dj{\hbox{d\kern-.33em\raise.80ex\hbox{-}
\raise-.80ex\hbox{\kern-.40em}}}
\usepackage{amssymb}
\usepackage{amscd,amssymb}
\usepackage[arrow,matrix]{xy}
\usepackage{graphicx}
\usepackage{epstopdf}

\usepackage{amscd,amsmath,amsthm,amssymb}

\theoremstyle{plain}

\newcommand{\sech}{\mbox{sech}}
\newcommand{\csch}{\mbox{csch}}

\newcommand{\bea}{\begin{eqnarray}}
\newcommand{\eea}{\end{eqnarray}}
\newcommand{\bes}{\begin{subequations}}
\newcommand{\ees}{\end{subequations}}

\theoremstyle{definition}

\begin{document}
\begin{frontmatter}
\title{Higher dimensional localized and periodic wave dynamics in a new integrable (2+1)-dimensional Kundu-Mukherjee-Naskar model}

\author[ss]{Sudhir Singh}
\author[am]{A. Mukherjee}
\author[ks1,ks2]{K. Sakkaravarthi\corref{kks}}\ead{ksakkaravarthi@gmail.com}
\author[ss]{K. Murugesan}

\address[ss]{Department of Mathematics, National Institute of Technology, Tiruchirappalli -- 620015, India}
\address[am]{Department of Physics, National University of Science and Technology, MISIS, Moscow, Russia}
\address[ks1]{Department of Physics, National Institute of Technology, Tiruchirappalli -- 620015, India}
\address[ks2]{Centre for Nonlinear Dynamics, School of Physics, Bharathidasan University, Tiruchirappalli -- 620024, India}

\cortext[kks]{Corresponding author}

\date{}

\begin{abstract}
In this article, a new integrable (2+1)-dimensional Kundu-Mukherjee-Naskar model which is a variant of the well known nonlinear Schr\"odinger equation is investigated. Bright-dark optical solitons along with periodic waves, complexiton and rational solutions are constructed by employing a generalized traveling wave analysis, Jacobian-elliptic function, Riccati equation and ansatz approach. Further, the dynamics of these bright/dark optical solitons and complexiton/periodic waves are studied by exploring the importance of arbitrary physical parameters with graphical demonstrations for a clear understanding. The obtained higher dimensional nonlinear wave solutions of this integrable system shall have useful applications in different physical systems including the dynamics of beam propagation in optical fibers, ion-acoustic waves in magnetized plasma and deep water oceanic rogue waves.
\end{abstract}


\end{frontmatter}
	
\section{Introduction}
Dynamics of nonlinear systems reveal several interesting phenomena which leads to a better understanding and influential applications in different fields of science, engineering and technology. This is due to the existence of various nonlinear coherent structures such as elliptic waves, solitons/solitary waves, rogue waves, breathers, shock waves, compactons, etc. associated with those systems. For the past few decades, the dynamics of such nonlinear waves have shown promising results in theoretical research as well as their corresponding experimental realization through which several applications have been demonstrated in diverse areas of physics like nonlinear optics, fluid dynamics, Bose-Einstein condensates, plasma physics, biophysics, etc. In general, these nonlinear waves possessing systems are modelled by nonlinear evolution (ordinary/partial differential) equations with various control parameters corresponding to the physical system under consideration. This ranges from the simple Kortewe-de Vries, sine-Gordon, nonlinear Schr\"odinger, Hirota, Burgers, Benjamin-Ono, Boussinesq, Camassa-Holm, Degasperis-Procesi, nonlinear $\sigma$ model, Heisenberg ferromagnet equations in one-dimension to Davey-Stewartson, Kadomtsev-Petviashvili, Navier-Stokes, Landau-Lifshitz model, Zakharov-Schulman, Gross-Pitaevskii equations in higher dimensions to name a few in different contexts of physics which appear as either integrable or non-integrable\cite{Yang-book}. The complications in the complete analysis of these and their related/extended version of nonlinear equations make them less well understood due to insufficient generalized mathematical tools. Hence, there is always a need for new integrable nonlinear systems due to their rich analytic beauty \cite{Yang-book,ML-book}. 

Integrable systems show wider variety of nonlinear coherent structures including the much celebrated solitons which attract huge attention not only due to their remarkable stability but also due to their intriguing collision dynamics. This enhances the understanding and consequences of different interactions in physics, such as the interaction between large amplitude waves along with wave-wave and wave-particle interactions. It is well known that the occurrence/existence of solitons is bound only to integrable nonlinear systems, while solitary waves (not solitons) appear in certain non-integrable as well as nearly-integrable models. Also, one has to note that a higher dimensional extension of an integrable one-dimensional model can become non-integrable and its dynamics will be restricted with a lesser number of free parameters \cite{Yang-book,ML-book,Kiv-book}. 
In view of the ever increasing interest on higher dimensional integrable nonlinear systems, the following new integrable (2+1) dimensional nonlinear evolution equation was proposed by Anjan Kundu, Abhik Mukherjee and Tapan Naskar to describe the dynamics two-dimensional oceanic rogue waves and two-dimensional ion-acoustic wave in a magnetized plasma \cite{Kundu-2014,Kundu-PoP-2015}:  
\bea \label{sec2eq1}
iq_t+q_{xy}+2iq(qq_x^*-q^*q_x)=0, 
\eea
where $q(x, y, t )$ is the envelope of nonlinear wave and asterisk corresponds to the complex conjugation. Here the first and second terms represent the temporal ($t$) evolution and the dispersion of the wave along two spatial ($x$ and $y$) dimensions. This equation was first derived in Ref. \cite{Kundu-2014} to model two-dimensional oceanic rogue waves and later it was named as Kundu-Mukherjee-Naskar (KMN) equation \cite{Kundu-PLA-2019,CNSNS-He-2016,Roman-2017,Optik-2019a,Optik-2019b,Optik-2019c}.  Importantly, it can be seen that the nonlinear terms are different from the conventional Kerr-type nonlinearity arising in the celebrated nonlinear Schr\"odinger (NLS) and other related/generalized models. Physically, it can be viewed as current-like nonlinearity arising from chirality as well as to study the phenomena of bending of light beams  \cite{CJP-2018}. Recently, the above KMN equation \eqref{sec2eq1} was shown to govern the dynamics of optical wave or soliton along an Erbium-doped coherently excited resonant waveguides \cite{CJP-2019}. Also, KMN equation \eqref{sec2eq1} possesses a rich dynamical characteristic similar to the standard NLS equation, and hence it can also be treated as a (2+1)-dimensional integrable generalization of the NLS equation. Due to these reasons, it has attracted considerable interest among the researchers including us for the present investigation, and several reports are available in the literature. 

It is important to highlight a few interesting observations made on KMN equation \eqref{sec2eq1}. Its integrability nature and rogue wave dynamics are reported in \cite{Kundu-2014}. Also, Hirota method was applied to obtain soliton and lump solutions of KMN model including its Lax pair, modulation instability and rogue waves in \cite{Kundu-PoP-2015, Kundu-PLA-2019}. Higher order rational/rogue wave solutions are also obtained with N-fold Darboux transformation method \cite{CNSNS-He-2016,Roman-2017} and a power series solution along with bright/dark soliton solutions have been studied by ansatz method \cite{Optik-2019a,Optik-2019b,Optik-2019c,CJP-2018,CJP-2019,MPLB-2018}. Still we can look for more general analytical solutions with different features resulting from more number of free/arbitrary parameters. In this aspect, one can study the periodic nonlinear waves and hyperbolic functions (elliptic waves) which give tremendous applications in nonlinear optics, plasma physics, hydrodynamics, etc. \cite{NLD-2DNLS-2017,PLA-2DNLS-2018,AMC-2018,PRE-2003} as they can explain many physical situations. One of the advantages of elliptic/periodic solutions is that they can be employed both for integrable as well as non-integrable evolution equations with different types of nonlinearities (for example focusing, defocussing and mixed type nonlinearities) \cite{Chow-OC-2003,PLA-2DNLS-2008,TK-PLA-2014,TK-CNSNS-2016,Optik-2KMN-2019,Optik-2KMN-2019a}. 

Based on the above, our objective of the present study is to obtain certain new type of periodic/elliptic/ hyperbolic solutions by applying the traveling wave ansatz and Jacobian elliptic solutions. Here we demonstrate the occurrence of bright and dark optical soliton solutions for specific choices of parameters in addition to the nonlinear periodic waves. Also, we have employed Ricatti equation method and ansatz function approach to construct complexiton and soliton solutions, where we have shown the importance of different wave parameters in their dynamics. Additionally, we attempt to find a doubly-localized dynamic rational solution of a forced KMN model. One of the interesting advantages of the present KMN model \eqref{sec2eq1} is that its nonlinear terms denoted by the derivatives provide additional features resulting into both bright as well as dark soliton solutions irrespective of its positive spatial dispersive effect/term comparing to the standard NLS equation which admits only bright (dark) soliton for focusing (defocusing) nonlinearity or with anamolous (normal) dispersion.

This article is structured in the following manner: In Sec. 2 a generalized traveling wave approach is adopted to obtain soliton, elliptic, periodic and unbounded solutions. Section 3 deals with complexiton solutions derived with the help of Riccati equation. In Sec. 4, an ansatz function approach is utilized to construct bright and dark solitons. Section 5 deals with the dynamic rational solutions while the final section is alloted for conclusions.  

\section{Generalized Traveling Wave Approach}
In this section, we derive exact analytical solutions of KMN equation (\ref{sec2eq1}) supporting different coherent structures. For this purpose, we apply the following wave transformation to Eq. \eqref{sec2eq1}
\bes \label{sec2eq2}
\bea
&q(x,y,t) = u(\varepsilon) e^{i \Phi }, \eea
where 
\bea & \varepsilon  = kx + ly -mt +a , \\
& \Phi  = px + ny - rt + b.
\eea
\ees
Here the function $u(\varepsilon)$ determines the shape of the nonlinear wave and the quantity $\Phi$ denotes the phase of the wave. Further in Eq. \eqref{sec2eq2} $k$, $l$, $m$, $a$, $p$, $n$, $r$ and $b$ are real parameters. 
Upon substitution of Eq. (\ref{sec2eq2}) into Eq. (\ref{sec2eq1}) and equating real/imaginary parts separately, we obtain 
\begin{equation} \label{sec2eq3}
\begin{aligned}
\text{Real Part }\qquad & : klu'' + 4pu^3 + (r-np)u=0, \\
\text{Imaginary Part }& : (m-kn-lp)u'=0.
\end{aligned}
\end{equation}
The above system of equations can be written in a convenient form as 
\bes\bea
\alpha u'' + \beta u^3 + \gamma u &=0 \label{sec2eq4}, \\
m - kn - lp &=0, \eea
 \label{sec2eq5} 
\ees
where $\alpha=kl$, $\beta =4 p$, and $\gamma = r-np$.  
Let us consider that the solution $u(\varepsilon)$ of equation (\ref{sec2eq4}) satisfies the Jacobian elliptic equation of the form 
\begin{equation} \label{sec2eq6}
u'^2 = h_0 + h_2 u^2 + h_4 u^4 , 
\end{equation}
where $h_0, h_2 $ and  $h_4$ are real constants to be determined. By differentiating the above equation (\ref{sec2eq6}) we obtain 
\begin{equation} \label{sec2eq7}
u'' = h_2 u + 2 h_4 u^3 .
\end{equation}
By comparing Eq. (\ref{sec2eq4}) with Eq. (\ref{sec2eq7}), and equating the coefficients of different powers of $u(\varepsilon)$ to zero, we obtain the following relations among the arbitrary real constants:
\begin{equation} \label{sec2eq8}
 h_4 = - \dfrac{\beta}{2 \alpha}, \quad h_2 = -\dfrac{\gamma}{\alpha}, \quad h_0= \mbox{arbitrary}.
\end{equation}
For various choices/combinations of the arbitrary constants appearing in the above equation, we can end up with different type of traveling wave solutions to the desired KMN equation \eqref{sec2eq1} which we are exploring in the following subsections.

\subsection{Bright Soliton Solution}
For the choice $h_0=0, h_2 >0$ and $h_4 <0$, Eq. (\ref{sec2eq6}) results into the following solution in the form of a hyperbolic function: 
\begin{equation} \label{sec2eq9}
u_1 (\varepsilon) = \sqrt{-\dfrac{h_2}{h_4} } \sech \left ( \sqrt{h_2} \varepsilon \right ). 
\end{equation}
Combining the above solution \eqref{sec2eq9} with Eq. \eqref{sec2eq5} and \eqref{sec2eq8}, we obtain the bright soliton solution of KMN model (\ref{sec2eq1}) as
\begin{equation} \label{sec2eq11}
q_1 =\sqrt{\dfrac{np-r}{2p}} \sech \left ( \sqrt{\dfrac{np-r}{kl}} ( kx+ly-(kn+lp)t+a)\right ) e^{i (px+ny-rt+b) }.
\end{equation}
The above bright optical soliton solution is characterized by seven arbitrary real parameters ($p,~n,~r$, $k,~l,~a$ and $b$) and admits bell-shaped profile structure with central peak with zero-background ($q= 0$ as $t\rightarrow \pm \infty$). Here the term $\sqrt{\dfrac{np-r}{2p}}$ represent the amplitude of the soliton while the factors $\frac{kn+lp}{k}$ and $\frac{kn+lp}{l}$ denote the velocity/speed of soliton during its propagation along $x$ and $y$ axis/direction, respectively. This shows that the soliton velocities are different in both spatial directions but with same amplitude. As described earlier, the exponential term corresponds to the phase which plays no crucial role in the propagation. For a better understanding, we have illustrated the spatiotemporal evolution of the bright soliton  \eqref{sec2eq11} in two different planes $x-t$ and $y-t$ in Fig. 1 for a particular choice of the parameters as given in the figure caption.
\begin{figure}[h] \label{fig1-sech}
	\begin{center} 
		\includegraphics[width=0.57\linewidth]{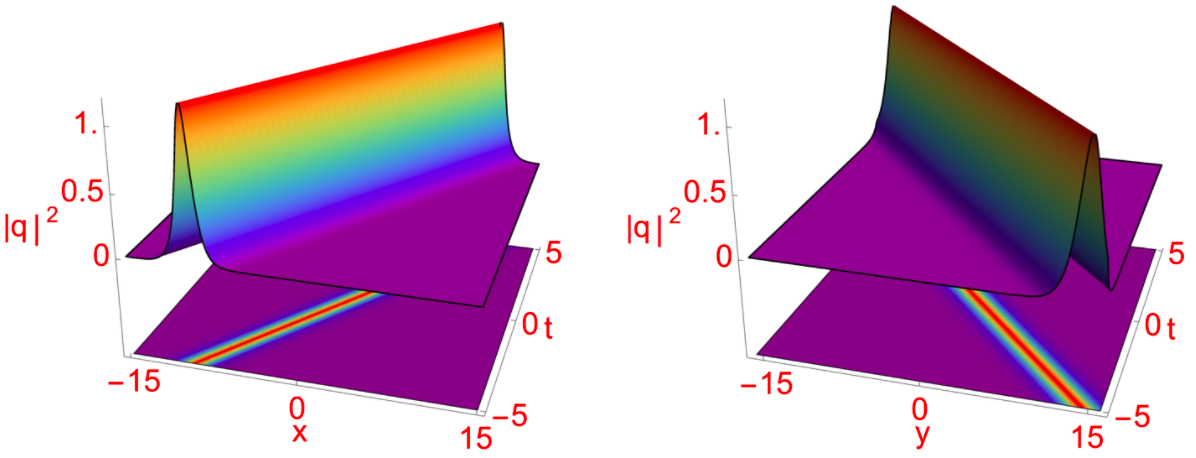}
		\caption{Evolution of bright optical soliton \eqref{sec2eq11} in the $x-t$ (left panel) and $y-t$ (right panel) planes for the parameter choice $r = 1$, $n = 1.5$, $p = -0.7$, $k = -1.05$, $l = 0.75$, $a = 0.2$ and $b = 0.4$ at $y=0.5$ and $x=0.5$, respectively, along with its intensity/density projection.}
	\end{center}\vspace{-0.52cm}
\end{figure}

\subsection{Dark Soliton Solution}
Opposite to the bright soliton solution, when $h_2 <0 $  and $h_4>0$ with $h_0 = \dfrac{h_2^2}{4h_4}$, Eq. (\ref{sec2eq6}) admits a tan-hyperbolic solution as given below. 
\begin{equation} \label{sec2eq12}
u_2 (\varepsilon) = \sqrt{ - \dfrac{h_2}{2 h_4} } \tanh \left ( \sqrt{- \dfrac{h_2}{2} } \varepsilon \right ).
\end{equation}
From this we obtain the exact solution of Eq. \eqref{sec2eq1} as 
\begin{equation} \label{sec2eq13}
q_2= \sqrt{\dfrac{np-r}{4p}} \tanh \left ( \sqrt{\dfrac{r-np}{2kl}}  ( kx+ly-(kn+lp)t+a)\right ) e^{i (px+ny-rt+b) }.
\end{equation}
The above `tanh' solution admit an inverted-bell shaped profile and is referred as dark soliton (localized dip or hole) on a constant (non-zero) background defined by the term $\sqrt{\dfrac{np-r}{4p}}$. Similar to the bright solitons, here also the velocity of dark solitons in along $x$ and $y$ directions are defined by $\frac{kn+lp}{k}$ and $\frac{kn+lp}{l}$ , respectively, which shows that the position of dark soliton at any given time along two planes are different during propagation. The space-time evolutions such dark soliton propagation are given in Fig. 2 for the parameter choice $r = -1.0$, $n = 1.5$, $p = 0.7$, $k = 2.0$, $l = -1.0$, $a = 0.2$ and $b = 0.4$. It is very clear from the Fig. 2 that the propagation of soliton has different velocity in both spatiotemporal dimensions.    
\begin{figure}[h]
	\begin{center}
		\includegraphics[width=0.57\linewidth]{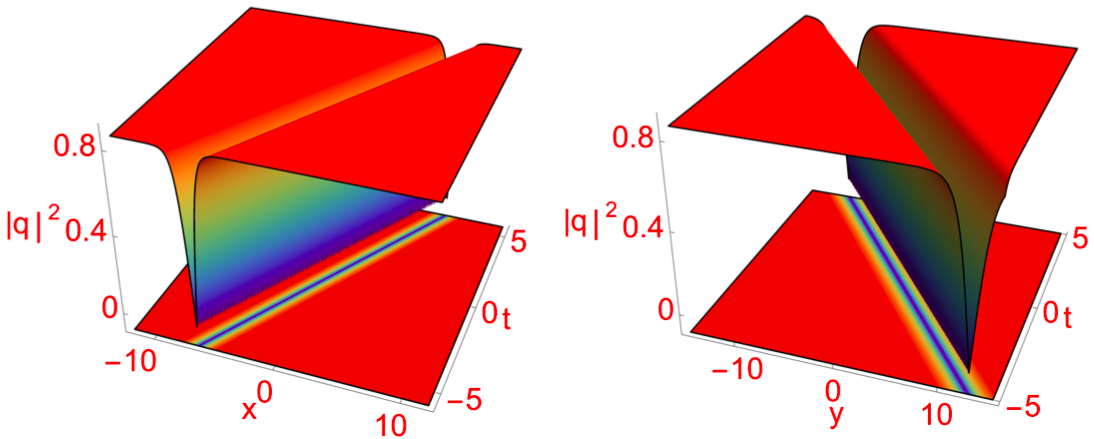}
		\caption{Evolution of dark optical soliton \eqref{sec2eq13} in the $x-t$ and $y-t$ planes respectively at $y=0.5$ and $x=0.5$ with its intensity dip as projection.}
	\end{center}
\end{figure}

As discussed in the introduction, it is evident from the above solutions/analyses that the present KMN model supports both bright and dark solitons that are due to the existence of mixed type nonlinearities proving the equivalent results from focusing and defocusing nonlinear effects, respectively. This kind of bright and dark soliton occurrence are not possible in other scalar NLS family of equations arising in different physical systems. This shows the physical significance of the considered system and shows novel dynamics compared to already reported results. So, this KMN equation can be treated/observed as one of the important higher dimensional version of NLS type models and can be derived to govern various physical systems which shall be helpful in understanding its multifaceted applications. 

\subsection{Hyperbolic Singular Solution}\label{singular}
In contrary to the bright and dark soliton solutions, the choice $h_0=0, h_2 >0$ and $h_4>0$ leads to singular or unbounded solution to Eq. (\ref{sec2eq6}) in the form of hyperbolic function
\begin{equation} \label{sec2eq14}
u_3(\varepsilon) = \sqrt{ \dfrac{h_2}{h_4}} \csch \left(\sqrt{h_2} \varepsilon \right ).
\end{equation}
This further results into a singular solution of KMN equation (\ref{sec2eq1}) as 
\begin{equation} \label{sec2eq15}
q_3= \sqrt{\dfrac{r-np}{2p}}~ \csch \left ( \sqrt{\dfrac{np-r}{kl}}( kx+ly-(kn+lp)t+a)\right ) e^{i (px+ny-rt+b) }.
\end{equation}
Needless to express in detail, the above described singular solution satisfying the required condition $h_2,h_4>0$ and $h_0=0$ is graphically demonstrated in Fig. 3 for the parameter choice $r = 1$, $n = 0.5$, $p = 0.7$, $k = 1.5$, $l = -0.75$, $a = 0.2$ and $b = 0.4$. Note here that the profile is localized in $x$ or $y$ and propagates over time $t$, but periodically exhibit excitations/singularity in the amplitude. \vspace{-0.4cm}
\begin{figure}[h]
	\begin{center}
		\includegraphics[width=0.8\linewidth]{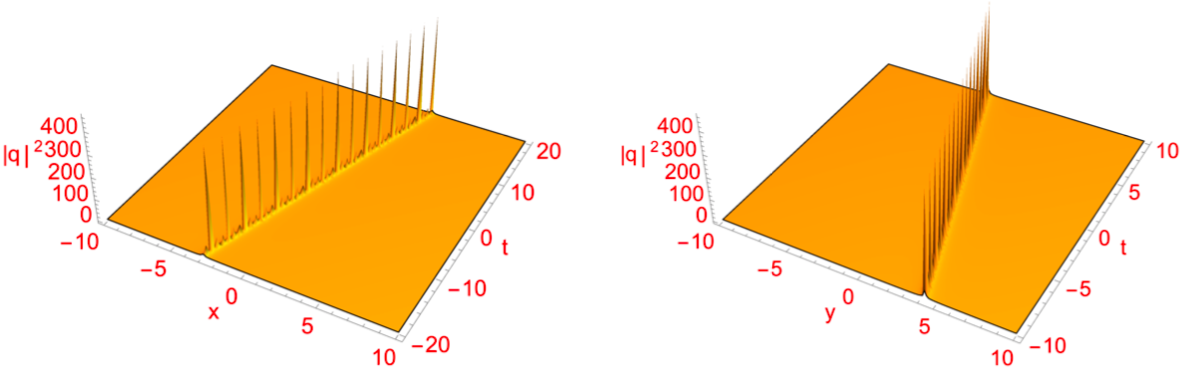}
		\caption{Localized periodic singular/unbounded structure  \eqref{sec2eq15} in the $x-t$ plane at $y=0.5$ and $y-t$ plane at $x=0.5$.}
	\end{center}
\end{figure}

\subsection{Trigonometric Unbounded Solutions}\label{unbound}
In contrary to the hyperbolic solutions listed in the above three cases, there exists another set of unbounded solutions to Eq. (\ref{sec2eq6}) and hence to KMN equation \eqref{sec2eq1} in the form of trigonometric functions for $h_4>0$ as given below.
\bes\bea
u_4 (\varepsilon) = \sqrt{- \dfrac{h_2}{h_4}} \sec \left (  \sqrt{-h_2} \varepsilon \right ), \qquad \text{ for } h_0 =0, h_2 <0,  \label{sec2eq16}\\ 
u_5 (\varepsilon) = \sqrt{\dfrac{h_2}{2h_4}} \tan \left (  \sqrt{ \dfrac{h_2}{2}} \varepsilon \right ), \qquad \text{ for } h_0 = \dfrac{h_2^2}{4h_4} , h_2 >0.  \label{sec2eq17}
\eea\ees
\begin{figure}[h]
	\begin{center}
		\includegraphics[width=0.8\linewidth]{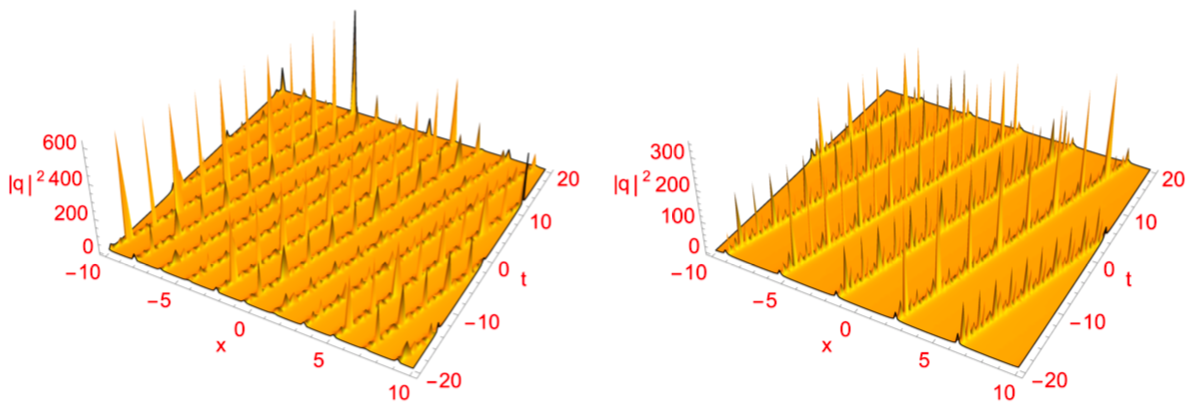}
		\caption{Singular/unbounded structures resulting for the trigonometric solutions \eqref{sec2eq19} and \eqref{sec2eq19a} in the $x-t$ plane for the choice $r = 1$, $n = 0.5$, $p = 0.7$, $k = 1.5$, $l = -0.75$, $a = 0.2$ and $b = 0.4$ at $y=0.5$.}
	\end{center}
\end{figure}
Thus we can construct two types of singular periodic traveling wave solutions to equation (\ref{sec2eq1}).
\bes\bea
q_4 &  = \sqrt{\dfrac{np-r}{2p}} \sec \left ( \sqrt{\dfrac{np-r}{kl}}( kx+ly-(kn+lp)t+a)\right )  e^{i (px+ny-rt+b) }, \label{sec2eq19}  \\
q_5 &  =\sqrt{\dfrac{r-np}{4p}} \tan  \left ( \sqrt{\dfrac{np-r}{2kl}}  ( kx+ly-(kn+lp)t+a)\right )  e^{i (px+ny-rt+b) } \label{sec2eq19a}.
\eea\ees 
The above trigonometric solutions evolve as singularities and their structures become unbounded and unlocalized in both $x$ and $y$, which is different from the localized singular structure presented in {\it Sec.} \ref{singular}. We have given the graphical demonstration of such singular solution profiles in Fig. 4 for completeness.

\subsection{Jacobi Elliptic, Periodic and Soliton Solutions}\label{jacobi}
In addition to the hyperbolic and trigonometric solutions, KMN model \eqref{sec2eq1} also exhibit three family of Jacobian elliptic function solutions for different choices of $h_0$, $h_2$ and $h_4$ given in \eqref{sec2eq8} which act as control parameters of the solution and they can be described as below. 
\bes\bea 
u_6 (\varepsilon) & =& \sqrt{- \dfrac{h_2m^2}{h_4 \alpha_1} }\,  cn \left ( \sqrt{ \dfrac{h_2}{\alpha_1}} \varepsilon , m \right ), \qquad \text{ for \quad} h_0 = \dfrac{h_2^2 m^2 (m^2-1) }{h_4 \alpha _1^2 }, h_2h_4 <0, \label{sec2eq20} \\
u_7 (\varepsilon) & =& \sqrt{- \dfrac{h_2 }{h_4 \alpha_2} }\,  dn \left ( \sqrt{ \dfrac{h_2}{\alpha_2}} \varepsilon , m \right ), \qquad\text{ for\quad } h_0 = \dfrac{h_2^2  (1-m^2) }{h_4 \alpha _2^2 }, h_2>0, h_4 <0, \label{sec2eq21} \\
u_8 (\varepsilon) & =& \sqrt{- \dfrac{h_2 m^2 }{h_4 \alpha_3} }\,  sn \left ( \sqrt{ \dfrac{ - h_2}{\alpha_3}} \varepsilon , m \right ), \qquad \text{ for\quad  } h_0 = \dfrac{h_2^2  m^2 }{h_4 \alpha _3^2 }, h_2<0, h_4 >0, \label{sec2eq22} 
\eea\ees 
where $\alpha_1 = 2m^2 -1 , \alpha_2 = 2-m^2$, $\alpha_3 = m^2 + 1 $ and $m$ is the elliptic modulus parameter. 
From the above forms, we can obtain the explicit `cnoidal', `dnoidal', and `snoidal' wave solutions of the KMN equation (\ref{sec2eq1}), respectively as 
\bes\bea 
q_6 &  =& \sqrt{ \dfrac{(np-r)m^2}{2p \alpha_1}} ~cn \left ( \sqrt{ \dfrac{np-r}{kl\alpha_1}  } ( kx+ly-(kn+lp)t+a),m\right)  e^{i (px+ny-rt+b)}, \label{sec2eq23} \\ 
q_7 &  =& \sqrt{\dfrac{np-r}{2p \alpha_2 } } ~dn \left ( \sqrt{ \dfrac{r+np}{kl \alpha_2}  } ( kx+ly-(kn+lp)t+a),m\right )  e^{i (px+ny-rt+b)}, \label{sec2eq24} \\
q_8 &  =& \sqrt{\dfrac{(np-r)m^2}{2p \alpha_3}} ~sn \left ( \sqrt{ - \left ( \dfrac{r+np}{k l \alpha_3}  \right ) } ( kx+ly-(kn+lp)t+a),m\right ) e^{i (px+ny-rt+b)}. \label{sec2eq25} 
\eea\ees 

To be precise, all the above elliptic function (cnoidal, dnoidal and snoidal) solutions  \eqref{sec2eq23}--\eqref{sec2eq25} admit both periodic and soliton/solitary wave solution for suitable elliptic modulus. Particularly, they exhibit periodic waves propagating along both spatial dimensions for elliptic modulus value $m=-1$. However for $m=1$, the `cnoidal' and `dnoidal' solutions exhibit bright solitons in zero-background, while the `snoidal' solution leads to a dark soliton structure on a non-zero constant background. Here also the velocity of periodic waves as well as the solitons depends on the arbitrary real parameters $r$, $n$, $p$, $k$, $l $, $a$ and $b$. For a better understanding and completeness, the evolution of such periodic waves and bright/dark solitons are illustrated in Figs. 5--7. 

\begin{figure}[h]
	\begin{center} 
		\includegraphics[width=0.98\linewidth]{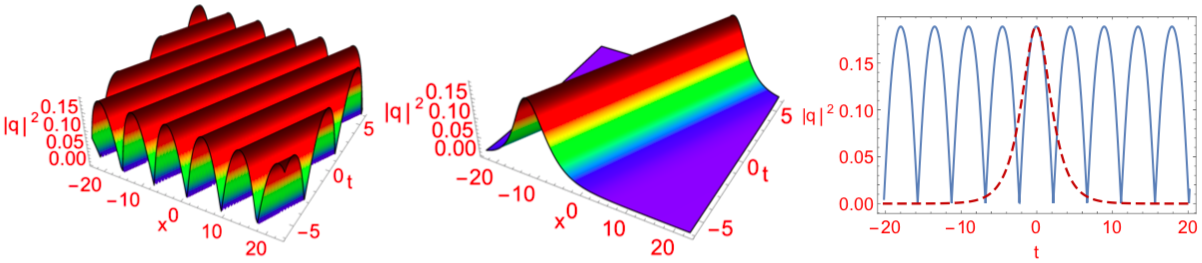}
		\caption{Evolution of periodic wave (left panel for $m=-1$) and bright soliton (middle panel for $m=1$) solution \eqref{sec2eq23} of KMN model in the $x-t$ plane for the parameter choice $r = 1$, $n = 1.5$, $p = 0.7$, $k = 1.5$, $l = 0.75$, $a = 0.2$ and $b = 0.4$ at $y=0.5$. Corresponding two-dimensional plot at $x=0.5$ and $y=0.5$ is shown in the right panel.}
	\end{center}  \label{cnoidal}
\end{figure} 
\begin{figure}[h] \label{dnoidal}
	\begin{center}
		\includegraphics[width=0.98\linewidth]{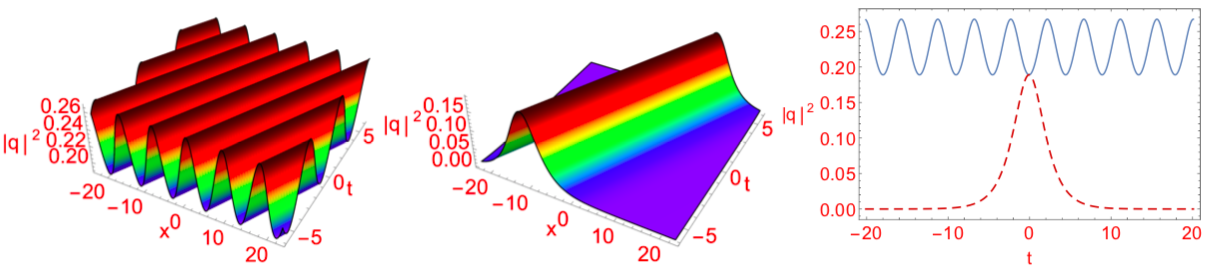}
		\caption{Periodic solution (left panel for $m=-1$) and bright soliton solution (middle panel for $m=1$) of KMN model given by \eqref{sec2eq24} in the $x-t$ plane for $r = 1$, $n = 1.5$, $p = 0.7$, $k = 1.5$, $l = 0.75$, $a = 0.2$ and $b = 0.4$ at $y=0.5$. Right panel shows their two-dimensional plot at $x=0.5$ and $y=0.5$.}
	\end{center}
\end{figure}
\begin{figure}[h] \label{snoidal}
	\begin{center}
		\includegraphics[width=0.98\linewidth]{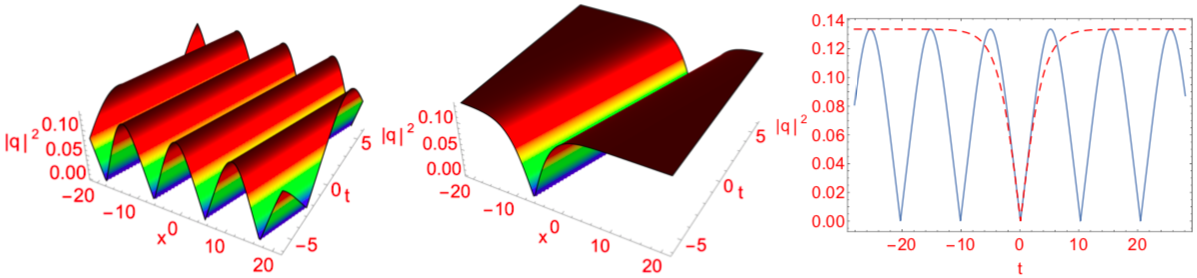}
		\caption{Periodic wave (left panel for $m=-1$) and dark soliton (middle panel for $m=1$) solution given by \eqref{sec2eq25} in the $x-t$ plane for $r = 1$, $n = 1.5$, $p = 0.7$, $k = -1.5$, $l = 0.75$, $a = 0.2$ and $b = 0.4$ at $y=0.5$. Right panel shows their two-dimensional structure at $x=0.5$ and $y=0.5$.}
	\end{center}
\end{figure}

\newpage
\section{Complexitons}
Complexitons are nothing but non-singular periodic solitons and they are new type of exact solutions admitted by certain nonlinear evolution equations. In this section, we obtain complexiton solutions by using the Riccati equation method \cite{PLA-2005}, which results into the form of periodic/trigonometric and hyperbolic function solutions. In order to compute such complexiton solutions, we use the following assumption:
\begin{equation} \label{sec3eq1}
q(x,y,t) = f(\xi) e^{i ( \alpha_1 x + \alpha_2 y + \beta t + \theta _0)},
\end{equation}
where $\xi=l_1x + l_2 y -vt + \theta_1 $ is a real-valued function in which $l_1$, $l_2$, $v$ and $\theta_0$ are real constants to be determined. Substituting Eq. (\ref{sec3eq1}) into the KMN equation (\ref{sec2eq1}), we get the following nonlinear ODE: 
\bea
& A_1 f'' + A_2 f' + A _3 f + A_4 f^3 =0, \label{sec3eq2}
\eea 
where $A_1 = - l_1 l_2$, $A_2  = i ( v - \alpha_2 l_1 - \alpha_1 l_2 )$, $A_3 = \beta + \alpha _1 \alpha _2$, and $A_4  = - 4 \alpha_1$. 
By balancing the linear term arising as the highest order derivative $f''$ and the nonlinear term $f^3$ from the above ODE, we can obtain the solution $f$ as
\begin{equation}
f= a_0 + a_1 \psi, \label{sec3eq4}
\end{equation}
where $\psi$ satisfies the Riccati equation $\psi ' = R + \psi ^2$, 
which has the following four set of solutions for two choices of $R$ \cite{PLA-2005}: 
\bes\bea 
\mbox{Choice 1: $R < 0$} \qquad &&\psi_1 = - \sqrt{- R} \,  \tanh(\sqrt{-R} \xi) , \qquad \psi_2 = - \sqrt{-R} \, \coth(\sqrt{-R}\xi )\,,  \label{sec3eq6}\\
\mbox{Choice 2: $R > 0$} \qquad &&\psi_3 = \sqrt{R} \, \tan(\sqrt{R}\, \xi), \qquad \qquad\psi_4 = - \sqrt{R} \, \cot(\sqrt{R}\, \xi). \label{sec3eq7}
\eea \label{sec3eq67}\ees 

\noindent Now, substituting Eq. (\ref{sec3eq4}) along with the solutions \eqref{sec3eq67} into Eq. (\ref{sec3eq2}), one can get relations among different parameters as
\begin{align}
a_0 = 0,  \quad
a_1 = \sqrt{\dfrac{- l_1 l_2}{2 \alpha_1}},  \quad
v = l_2 \alpha _1 + l_1 \alpha_2,  \quad
\beta  = 2 R l_1 l_2 - \alpha_1 \alpha _2. \label{sec3eq8}
\end{align}
\begin{figure}[h]
	\begin{center} 
		\includegraphics[width=0.67\linewidth]{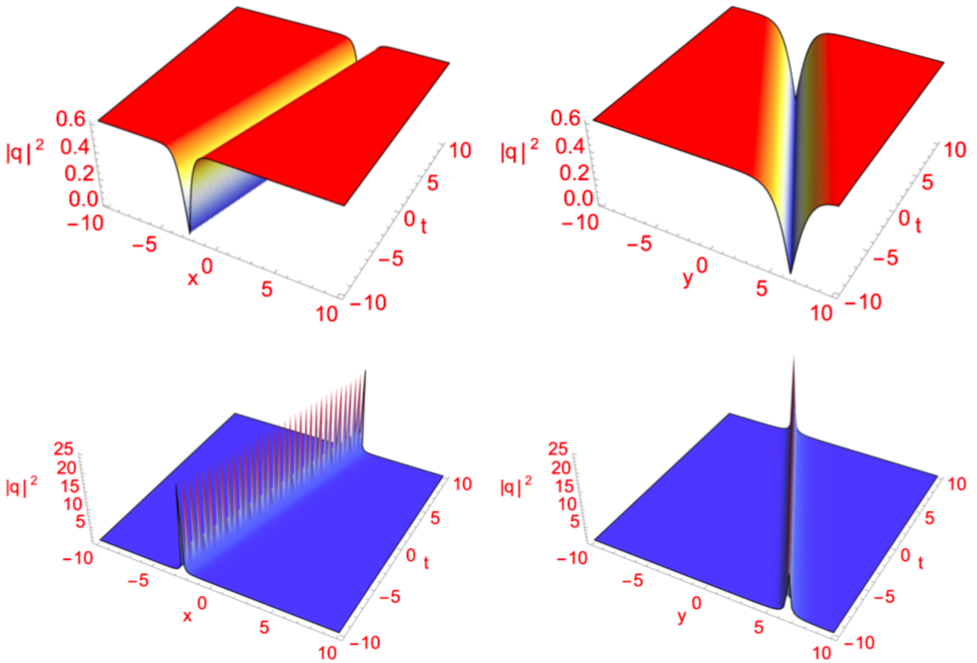}
		\caption{Complexiton of KMN system \eqref{sec2eq1} exhibiting dark  soliton \eqref{sec3eq9a} and localized sigular excitations \eqref{sec3eq9b} in the $x-t$ ($y-t$) plane for $y=0.5$ ($x=0.5$) with other parameters as $l_1= 1.5$, $l_2 = -0.75$, $\alpha_1=1.5$, $\alpha_2=1.0$, $\theta_0 = 0.4$, $\theta_1 =0.2$ and $R=-1$.}
	\end{center}\vspace{-0.42cm}
\end{figure}
Hence from the above, the resultant hyperbolic and trigonometric traveling wave solutions of KMN equation (\ref{sec2eq1}) can be obtained through Eq. (\ref{sec3eq1}) as below.
\bes\bea 
&&q_1 = \sqrt{\dfrac{- l_1 l_2 }{2 \alpha_1} } \left[- \sqrt{- R} \, \tanh\left(  \sqrt{-R} \, (l_1 x + l_2 y - vt + \theta _1 )\right) \right] e^{i ( \alpha_1 x + \alpha_2 y + \beta t  + \theta _0)}, \label{sec3eq9a} \\
&&q_2 = \sqrt{\dfrac{- l_1 l_2 }{2 \alpha_1} } \left[- \sqrt{ - R} \, \coth\left(  \sqrt{-R} \, (l_1 x + l_2 y - vt + \theta _1 ) \right)\right] e^{i ( \alpha_1 x + \alpha_2 y + \beta t  + \theta _0)}, \label{sec3eq9b}\\
&& q_3 = \sqrt{\dfrac{- l_1 l_2 }{2 \alpha_1} } \left[ \sqrt{R} \, \tan \left(  \sqrt{R} \, (l_1 x + l_2 y - vt + \theta _1 ) \right )\right] e^{i ( \alpha_1 x + \alpha_2 y + \beta t  + \theta _0)}, \label{sec3eq9c} \\
&&q_4= \sqrt{\dfrac{- l_1 l_2 }{2 \alpha_1} } \left[- \sqrt{R} \, \cot \left( \sqrt{R} \, (l_1 x + l_2 y - vt + \theta _1 ) \right )\right] e^{i ( \alpha_1 x + \alpha_2 y + \beta t  + \theta _0)}. \label{sec3eq9d}
\eea \label{sec3eq9ad} \ees 
For different choices of arbitrary parameters in Eq. \eqref{sec3eq9ad}, we can evidence bounded regular as well as unbounded singular solutions. Particularly, we obtain dark solitons resulting for `tanh' solution, while the other solutions lead to localized/unlocalized singular structures. We have illustrated such complexiton solutions exhibiting dark soliton and singular structures in Figs. 8 and 9 for appropriate parameters. To be precise, the bounded regular dark soliton \eqref{sec3eq9a} possesses an amplitude defined as $ \sqrt{\dfrac{- l_1 l_2 }{2 \alpha_1} } $ while the velocity takes the form $v/l_1$ and $v/l_2$ during its propagation along $x-t$ and $y-t$ spatiotemporal dimensions, respectively. 
\begin{figure}[h]
	\begin{center} 
		\includegraphics[width=0.7\linewidth]{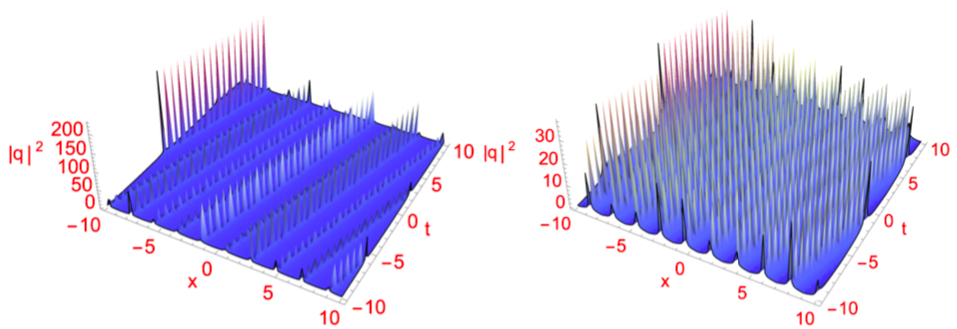}
		\caption{Unbounded singular structure of complexiton solution given by \eqref{sec3eq9c} and \eqref{sec3eq9d}, respectively in the $x-t$ plane for $y=0.5$ with other parameters as $l_1= 1.5$, $l_2 = -0.75$, $\alpha_1=1.5$, $\alpha_2=1.0$, $\theta_0 = 0.4$, $\theta_1 =0.2$ and $R=1$. Similar unbounded structures appear in $y-t$ plane also.}
	\end{center} \vspace{-0.752cm}
\end{figure}

\section{Simplified Ansatz Approach}

Though this KMN model is integrable and more delicate methods such as standard Hirota bilinearization, Darboux transformation, etc. can be applied to find its soliton solutions, here we utilize a simplified approach to compare/analyze its limitations/advantages in exploring the integrable nonlinear evolution equations. In view of this, we apply an ansatz approach to find the soliton solutions of KMN equation (\ref{sec2eq1}) and compare them with already obtained solutions in the previous sections. For this purpose, we consider an ansatz 
\begin{equation}
q= u(x,y,t) e^{i(\theta x + \psi y+ \phi t+\epsilon_0)}, \label{eq1}
\end{equation}
where $u(x,y,t)$ is a real function while $\theta$, $\psi$, $\phi$ are real parameters. By adopting the above form for $q$ \eqref{eq1}, the complex KMN equation (\ref{sec2eq1}) can be transformed into the following two real equations after separating the real and imaginary parts:
\bes\bea
&u_{xy}- \phi u-\psi \theta u + 4 \theta u^3 =0,\\
&u_t + \theta u_y+\psi u_x=0. 
\eea  \label{eq3}\ees 
Solving the above equations \eqref{eq3} can end up with various solutions for $u(x,y,t)$. Here we concentrate on the hyperbolic solutions to describe bright/dark solitons and briefly analyze their evolution. 

\subsection{Bright soliton solution} 
Let us suppose that solution of equation (\ref{eq3}) can be written in a rational form of two real functions $g(x,y,t)$ and $f(x,y,t)$ as 
\begin{equation}
u=\dfrac{g}{f}, \qquad f=1+e^{2 \xi }, \quad g=2u_0 e^\xi, \label{eq4}
\end{equation}
with $\xi = kx + ly + mt+ \varepsilon _0 $, in which $u_0$, $k$, $l$, $m$ and $\varepsilon _0 $ are arbitrary real constants to be determined. Now substituting Eq. \eqref{eq4} into (\ref{eq3}) solving the resultant equation arising as the coefficients at different orders of $e^\xi$, we get the bright soliton solution of KMN equation (\ref{sec2eq1}) as 
\bes\bea 
q=u_0 e^{i(\theta x + \psi y+ \phi t+\epsilon_0)}~ \sech (kx+ly+mt+\varepsilon_0), \label{eq5}
\eea
along with certain conditions on constant parameters  
\bea
2 \theta u_0^2 ={k l},\quad \phi=k l-\psi \theta, \quad m+k \psi +l  \theta=0.
\eea \label{bright}\ees 
The bright soliton solution \eqref{bright} is characterized by only five arbitrary parameters. Here we can find that the soliton velocity is $-m/k$ and $-m/l$ along the spatiotemporal dimensions $x-t$ and $y-t$, respectively, while its amplitude is purely depends on the parameter $u_0$. A schematic representation for the evolution of above bright soliton \eqref{bright} is given in Fig. 10 for the choice of arbitrary parameters $k = -1.05$, $l = 0.5$, $\psi = 0.5$, $\varepsilon_0 = 0.2$, $\epsilon_0=0$, and $u_0 = 1.0$. 
\begin{figure}[h]
	\begin{center} 
		\includegraphics[width=0.75\linewidth]{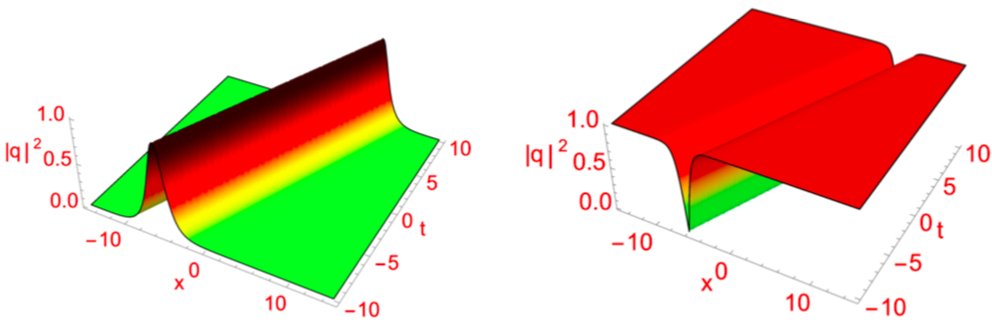}
		\caption{Bright and dark soliton solutions of KMN system given by \eqref{bright} and \eqref{dark} in the $x-t$ plane for $y=0.5$.}
	\end{center} \label{ansatz-soliton-fig}\vspace{-0.752cm}
\end{figure}

\subsection{Dark soliton solution} 
In addition to the bright soliton solution, when we choose the form of $g$ and $f$ as 
\bea u=\dfrac{g}{f}, \qquad  f=1+e^{2 \xi }, \quad g=u_0(-1+e^{2 \xi }).\eea 
Equation \eqref{eq3} results into the following relations
\bes\bea 
2 \theta u_0 ^2  = {k l}, \quad \phi = 2kl-\psi \theta = 0, \quad m+k \psi +l \theta =0.
\eea
Then, the resultant dark soliton solution can be expressed as
\bea
q=u_0  \tanh (lx+my+nt+\varepsilon_0)e^{i(\theta x + \psi y+ \phi t+\epsilon_0)}. \label{eq6}
\eea \label{dark} \ees 
It is very clear that $|q|_{t \to \pm \infty} \longrightarrow u_0 $ and the solutions are characterized by six arbitrary parameters, where the soliton velocity is represented by $-m/k$ along $x-t$ plane and $-m/l$ along $y-t$ plane. A simple graphical demonstration of the dark soliton is given in Fig. 10 for $k = -1.05$, $l = 0.5$, $\psi = 0.5$, $\varepsilon_0 = 0.2$, $\epsilon_0=0$, and $u_0 = 1.0$. 

In view of the current objective, the periodic and soliton solutions obtained in this work are characterized by more number of arbitrary parameters than that of already reported results in the literature. Among the solutions presented in the present work itself, the bright/dark soliton and periodic wave solutions obtained by the generalized traveling wave approach in Sec. 2 \& 3 have more number of arbitrary parameters than that of complexitons (Sec. 4) and simplified ansatz approach (Sec. 5) which shall provide additional freedom in defining the solitons appropriately and to manipulate their dynamics based on our requirement for suitable applications.


\section{Rational Solutions}
In addition to the above discussed soliton and periodic solutions, in this section we examine the rational solution of KMN equation. Rational solutions are a branch of doubly-localized nonlinear wave structures with certain free/arbitrary parameters which define the nature of their evolution and dynamics. In one hand, they are closely related/resemble to rogue waves as well as peregrine solitons. On the other hand, in higher dimensional systems these rational solutions have strong conection with lump solutions \cite{CNSNS-He-2016}. Here one can also investigate the occurrence and dynamics of rational solution to the given system by considering the copropagation/coexistence of solitons/solitary waves which have attracted much attention in recent years. However, here we restrict our study to explore only the rational solution of KMN equation with an external time-dependent forcing. In fact, without the forcing term we shall have the presence of only static rational solution. In order to understand the features of a dynamical rational solution, we consider the following form of KMN model by including a forcing factor, which can also be called as modified KMN equation:
\bea i u_t + u_{xy} + 2 i u (u u^*_x - u^* u_x) = f(t) u. \label{roteq1}\eea
In the above equation $f(t)$ represents the external forcing term which can provide time dependent (dynamical) features of rational solution instead of a limited steady state (static) behaviour. Our aim in this section is to provide an infinite number of dynamic solution which can be obtained from a static solution through the following rotational transformation:
\bea 
u(x,y,t)=q(x,y,t) e^{i\int f(t))dt}. \label{roteq2}
\eea
\begin{figure}[h] 
	\begin{center} 
		\includegraphics[width=0.98\linewidth]{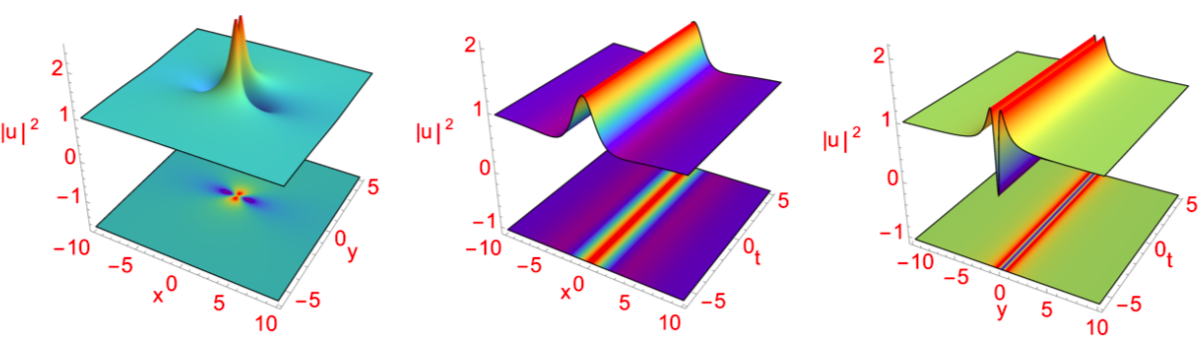}
		\caption{Doubly localized rational (lump) solution of forced KMN model \eqref{roteq1} for $f(t)=t^2+2t-3$, $a=0.5$ and $c=0.4$. From left to right the profile structure in $x-y$, $x-t$ and $y-t$ planes for a fixed $t=1.5$, $y=0.5$ and $x=0.5$, respectively. The modulus of the rational solution do not show any appreciable change over time.} 
	\end{center} \label{rational-fig}
\end{figure}
The above equation reduces \eqref{roteq1} into a standard KMN model \eqref{sec2eq1}. 
From the static rational solution corresponding to Eq. \eqref{sec2eq1} given in \cite{Kundu-2014,Kundu-PoP-2015} and with the help of rotational transformation \eqref{roteq2} we can construct a dynamical rational solution for arbitrary/different choices of $f(t)$ in the form given below.
\begin{equation}
u(x,y,t) = \left ( \dfrac{1 - 4 iy }{ax^2 + 4 y^2+c} -1\right ) e ^{4iy - i \int f(t) dt }. \label{rateq*}
\end{equation}
Here $a$ and $c$ are two arbitrary real constants which defines the nature of static lump/rational solution, while $f(t) \in C^0$ is a class of arbitrary continuous real function of time from which one can extract an infinitely many solutions of the given forced KMN model. 
\begin{figure}[h]
	\begin{center} 
		\includegraphics[width=0.7\linewidth]{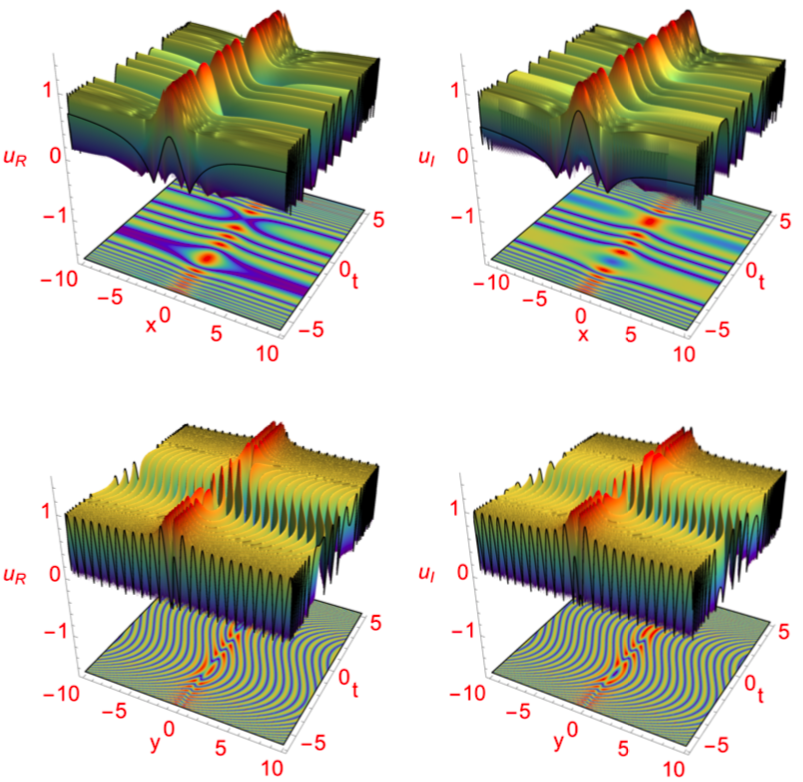}
		\caption{Time varying pattern of rational solution corresponding to the real/imaginary components in $x-t$ plane (top panel) and $y-t$ for $y=0.5$ plane (bottom panel) for $x=0.5$.} 
	\end{center} \label{rational-fig}
\end{figure}

Thorough study on the above dynamical rational solutions does not express any real observable effect as the time-dependency lies as a phase factor which plays very less or no role in real dynamics. This rational solution in $x-t$ and $y-t$ planes takes structure similar to that of peregrine solitons \cite{PNAS2019}. However, this shows significant observable effects if one investigate the real and imaginary components separately that depicts the occurrence of rational wave structure on a periodically modulated/oscillating background due to the time-varying external forcing factor. By suitably tuning the arbitrary parameters $a$ and $c$ one can control the steepness, inclination and amplitude of rational/rogue wave in addition to an appropriate forcing function. To elucidate the understanding, we have graphically demonstrated such rational wave structure in Figs. 11 and 12. Importantly, the rational solution exhibit oscillations in its amplitude and background for a given choice of external forcing $f(t)$.  

\section{Conclusions}
To conclude, we have constructed bright/dark optical soliton, complexiton and periodic solutions of the integrable (2+1)-dimensional Kundu-Mukherjee-Naskar equation by adopting Jacobian-elliptic function, Riccati equation and ansatz methods and analyzed their evolution for obtained arbitrary parameters. 
Interestingly, the present KMN model provides a nonlinear effect which can results from the mixture of both focusing and defocusing nonlinearities with spatial derivatives and this has proved the admittance of both bright and dark soliton solutions irrespective of the same type of dispersion. Additionally, a dynamical rational solution of a modified KMN model with an external forcing has been studied which attracts considerable interest in recent years. Consequently, we have shown that the propagation dynamics of these solitons having more number of arbitrary parameters can find added advantage over the solutions reported earlier in oceanic waves, nonlinear optical waves in coherently excited resonant waveguides, etc. As a further study, the present investigation can be extended to any other integrable systems. The present results will be an important contribution on higher dimensional nonlinear waves associated arising under different physical contexts. Also, an interesting future outlook is to explore the collision dynamics of multiple solitons which can be studied by constructing exact multi-bright and -dark soliton solutions of the KMN model and can look for possible non-trivial interaction scenario. Works are in progress along these directions and the results will be reported separately.
\vspace{-0.32cm}
\section*{Acknowledgments} \vspace{-0.2cm}
One of the authors SS would like to thank MHRD and National Institute of Technology, Tiruchirappalli, India for financial support through institute fellowship. KS was supported by DST-SERB National Post-Doctoral Fellowship (File No. PDF/2016/000547). 
 

\end{document}